\documentclass[times,astrosymb,twocolumn]{aastex631}

\usepackage{natbib}

\usepackage{hyperref}

\graphicspath{{./}{figures/}}
\usepackage{graphics,graphicx}
\usepackage{footnote}
\usepackage{aligned-overset}

\newcommand{\uat}[2]{\href{http://astrothesaurus.org/uat/#2}{#1 (#2)}}
\newcommand{\shortname}{J0100+2802}
\newcommand{\longname}{SDSS J010013.02+280225.8}

\received{August 2, 2021}
\revised{October 18, 2021}
\accepted{November 8, 2021}
\submitjournal{The Astrophysical Journal Letters}

\shorttitle{J0100+2802 in X-Rays -- Is It Lensed?}
\shortauthors{Connor et al.}

\graphicspath{{./}{figures/}}

\begin{document}

\title{X-Ray Evidence Against the Hypothesis that the Hyper-Luminous $z=6.3$ Quasar J0100+2802 is Lensed}

\correspondingauthor{Thomas Connor}
\email{thomas.p.connor@jpl.nasa.gov}

\author[0000-0002-7898-7664]{Thomas Connor}
\altaffiliation{NPP Fellow}
\affiliation{Jet Propulsion Laboratory, California Institute of Technology, 4800 Oak Grove Drive, Pasadena, CA 91109, USA}

\author[0000-0003-2686-9241]{Daniel Stern}
\affiliation{Jet Propulsion Laboratory, California Institute of Technology, 4800 Oak Grove Drive, Pasadena, CA 91109, USA}

\author[0000-0002-2931-7824]{Eduardo Ba\~nados}
\affiliation{Max Planck Institute for Astronomy, K\"onigstuhl 17, 69117 Heidelberg, Germany}

\author[0000-0002-5941-5214]{Chiara Mazzucchelli}
\affiliation{European Southern Observatory, Alonso de Cordova 3107, Vitacura, Region Metropolitana, Chile}

\begin{abstract}

The $z=6.327$ quasar SDSS J010013.02+280225.8 (hereafter J0100+2802) is believed to be powered by a black hole more massive than $10^{10}\ {\rm M}_\odot$, making it the most massive black hole known in the first billion years of the Universe. 
However, recent high-resolution ALMA imaging shows four structures at the location of this quasar, potentially implying that it is lensed with a magnification of $\mu\sim450$ and thus its black hole is significantly less massive. Furthermore, for the underlying distribution of magnifications of $z\gtrsim6$ quasars to produce such an extreme value, theoretical models predict that a larger number of quasars in this epoch should be lensed, implying further overestimates of early black hole masses. 
To provide an independent constraint on the possibility that J0100+2802 is lensed, we re-analyzed archival \textit{XMM-Newton} observations of the quasar and compared the expected ratios of X-ray luminosity to rest-frame UV and IR luminosities. For both cases, J0100+2802's X-ray flux is consistent with the no-lensing scenario; while this could be explained by J0100+2802 being X-ray faint, we find it does not have the X-ray or optical spectral features expected for an X-ray faint quasar. Finally, we compare the overall distribution of X-ray fluxes for known, typical $z\gtrsim6$ quasars.  We find a $3\sigma$ tension between the observed and predicted X-ray-to-UV flux ratios when adopting the magnification probability distribution required to produce a $\mu=450$ quasar.

\end{abstract}


\section{Introduction} \label{sec:intro}

The growing census of quasars known in the first billion years of the Universe ($z\gtrsim6$) has led to significant insights into the early universe, from how supermassive black holes (SMBHs) grow and evolve to the role that active galactic nuclei (AGN) play in processing the environments around the first massive galaxies. To date, over 200 quasars are known at $z>6$ \citep[e.g.,][]{2016ApJS..227...11B}; the most distant of these, at $z=7.64$, has mass $M_{\rm BH}=1.6\times10^9\ {\rm M}_\odot$, comparable to the most massive black holes in the local Universe \citep{2021ApJ...907L...1W}. These results are challenging to black hole formation models as, assuming Eddington-limited growth, a seed black hole formed 100 Myr after the Big Bang would require an initial mass greater than $10^4\ {\rm M}_\odot$ to produce the $z=7.64$ quasar.

\begin{figure*}
\centering
\vspace{5mm}
   \includegraphics{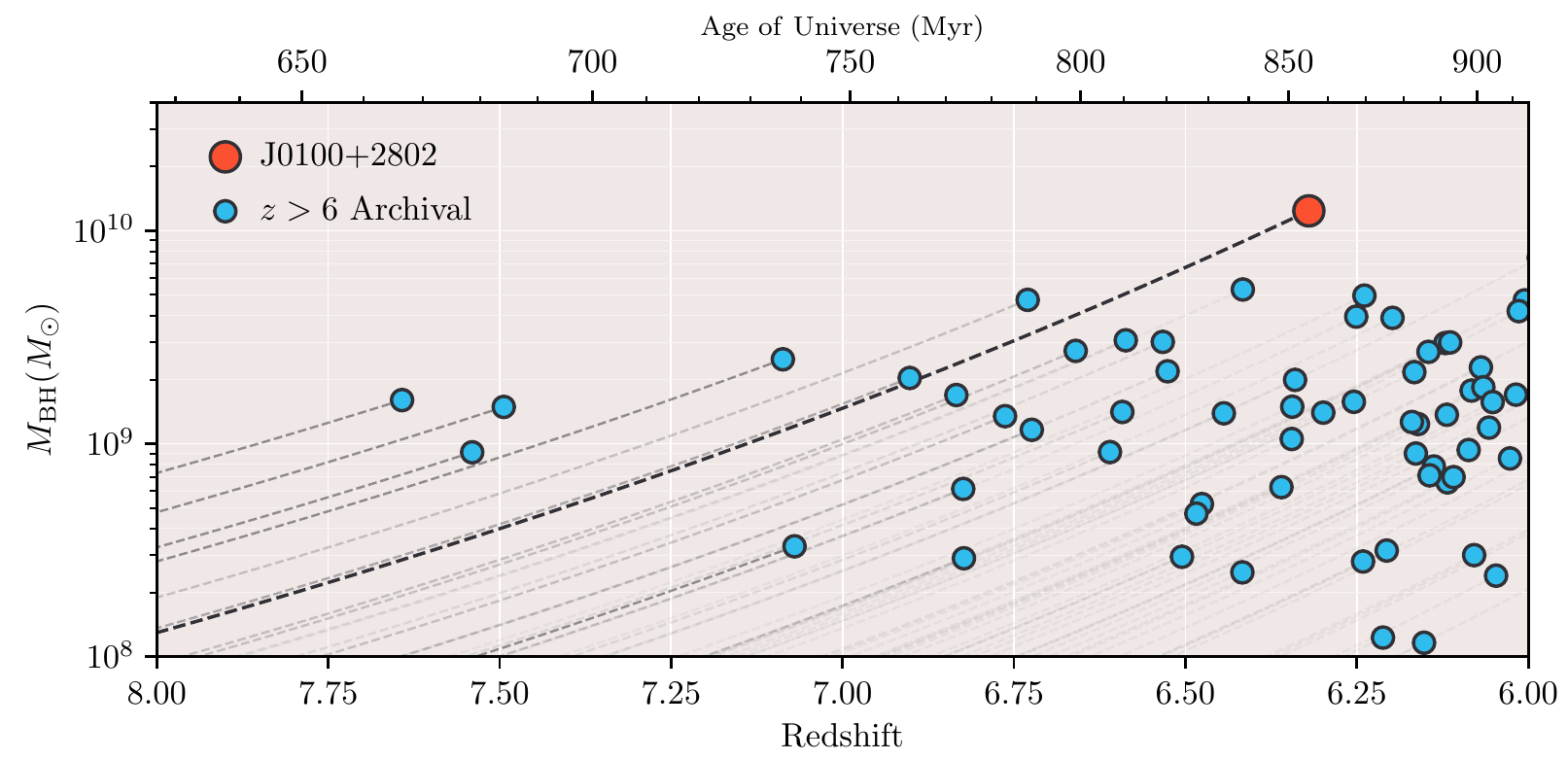}
      \caption{\small Mass and redshift of \shortname\ (red), in comparison to a near-complete sample of $z>6$ quasars currently known with robust, \ion{Mg}{2}-derived black hole mass estimates (blue). For all quasars, a mass growth track is shown with dashed lines, depicting the expected mass of the black hole at earlier times, assuming constant Eddington-limited growth with an efficiency of 0.1. While \shortname\ does not set the most stringent constraints on seed masses, it does nevertheless require significant, sustained growth through the entire Epoch of Reionization.
      }\label{fig:mass_distribution}
\end{figure*}

Chief among the high-redshift quasars is \longname\ (hereafter \shortname), a hyper-luminous ($L_{\rm bol}>10^{48}\ {\rm erg}\ {\rm s}^{-1}$), $z\sim6.3$ quasar with a \ion{Mg}{2}-derived mass of $M_{\rm BH} \sim 1.2 \times 10^{10}\ {\rm M}_\odot$ \citep{2015Natur.518..512W}. As shown in Figure \ref{fig:mass_distribution}, it is, by far, the most massive black hole known at $z\gtrsim6$. While, assuming Eddington-limited growth, \shortname\ does not set the most stringent limits on the seeds of early black holes, its significant mass does nevertheless require rapid growth to have been occurring from seed formation to 850 Myr after the Big Bang. Based on imaging with the Atacama Large Millimeter/submillimeter Array (ALMA), \citet{2019ApJ...880....2W} measured a kinematic mass for \shortname's host galaxy of $M_{dyn}\sim 3.3 \times 10^{10}\ {\rm M}_\odot$. This value is comparable to other high-z quasar host galaxies, although the implied black hole to dynamical mass fraction of ${\sim}35\%$ is the largest value observed at $z\gtrsim6$ \citep[e.g.,][]{2015ApJ...801..123W}. 

Recently, \citet{2020ApJ...891...64F} re-analyzed the ALMA observations of \shortname. In contrast to \citet{2019ApJ...880....2W}, \citet{2020ApJ...891...64F} generated higher-resolution images by utilizing Briggs weighting with a robust parameter of 0.2, resulting in a synthesized beam size of $0\farcs21 \times 0\farcs09$. In the new, higher-resolution maps, \citet{2020ApJ...891...64F} resolved \shortname\ into a compact quadruple system within a radius of ${\sim}0\farcs2$. They proposed two explanations for this: either there are multiple, dusty star-forming regions in \shortname's host galaxy or the quasar has been lensed into a quad. They argued for the latter on account of emission and absorption features from a foreground $z=2.33$ object in the quasar's spectrum. The mass model from \textit{HST} imaging implies a magnification of $\mu{\sim}450$. Such a magnification is extreme, with the closest analog coming from \citet{2018arXiv180705434G}, who report a $z=2.5$ quad-lensed quasar that is magnified by a factor of $\mu \sim 100$. This value also dwarfs that of UHS J0439+1634, the only reionization-era quasar known to be lensed, which has a magnification of $\mu {\sim} 50$ \citep{2019ApJ...870L..11F}. 

If \shortname\ is lensed, a magnification of $\mu{\sim}450$ would be significant; as the black hole mass estimate is based partially on the quasar's luminosity, this would imply $M_{\rm BH}<10^{9}\ {\rm M}_\odot$. And not only does $\mu{\sim}450$ for \shortname\ remove one of the strongest pieces of evidence for significant mass growth in late-stage reionization, but it could also imply broader trends; \citet{2020ApJ...889...52P} argue that, for a $\mu{\sim}450$ quasar to exist, the expected distribution of magnification values would require that more of the $z>6$ quasar population is lensed by $\mu \gtrsim 10$.

So far, the strongest rebuttal to the proposed lensing explanation for \shortname\ comes from \citet{2020ApJ...904L..32D}. They point to the proximity zone of the quasar, the region around the AGN that has been substantially overionized, leading to a deficit in absorption observed in the spectrum blueward of Ly$\alpha$. \citet{2017ApJ...840...24E} originally noted that \shortname\ has an unusually small proximity zone ($R_p=7.1$ proper Mpc, or a luminosity-corrected $R_{p,corr}=3.1$ proper Mpc). Such a small zone size could imply a short quasar lifetime ($t < 10^5$ yr) -- compared to values of $t>10^6$ yr seen at this epoch in other proximity zone measurements \citep{2017ApJ...840...24E} and from jet lifetimes \citep{2021ApJ...911..120C} -- or that the intrinsic quasar luminosity is less than what is observed.

In the case of \shortname, \citet{2020ApJ...904L..32D} simulated the evolution of 200 massive halos, each with a grid of assumed intrinsic quasar luminosities. The resultant simulated spectra that most closely matched the observed spectrum of \shortname\ came from quasars with a short lifetime ($\lesssim10^{5}$ yr) and the observed luminosity of the quasar, implying that it is not magnified by lensing. \citet{2020ApJ...904L..32D} also demonstrated the feasibility of this technique by recovering the known magnification of UHS J0439+1634 \citep{2019ApJ...870L..11F}.

To provide an observational counterpart to the simulation-based results of \citet{2020ApJ...904L..32D}, we turn to the X-ray properties of \shortname. Rest-frame X-ray, UV, and IR emission are generated through different processes and from different regions in an AGN. These regions will behave differently as AGN luminosity increases -- particularly the X-ray producing corona, which saturates at high luminosities. As such, the flux ratios in these bands provide a marker of the intrinsic luminosity \citep[e.g.,][]{2017A&A...608A..51M}, and can therefore be used to constrain the magnification of \shortname. Here, we re-analyze the X-ray observations of \shortname, focusing specifically on how consistent its relative luminosities are with an interpretation that it is magnified by $\mu{\sim}450$. Throughout this work we adopt a redshift of $z=6.327$, based on the [\ion{C}{2}] measurements of \citet{2019ApJ...880....2W}, and we assume a flat cosmology with $H_0 = 70\,\textrm{km\,s}^{-1}\,\textrm{Mpc}^{-1}$, $\Omega_M = 0.3$, and $\Omega_\Lambda = 0.7$.

\section{X-ray Observations}

\begin{figure*}
\centering
\vspace{5mm}
   \includegraphics{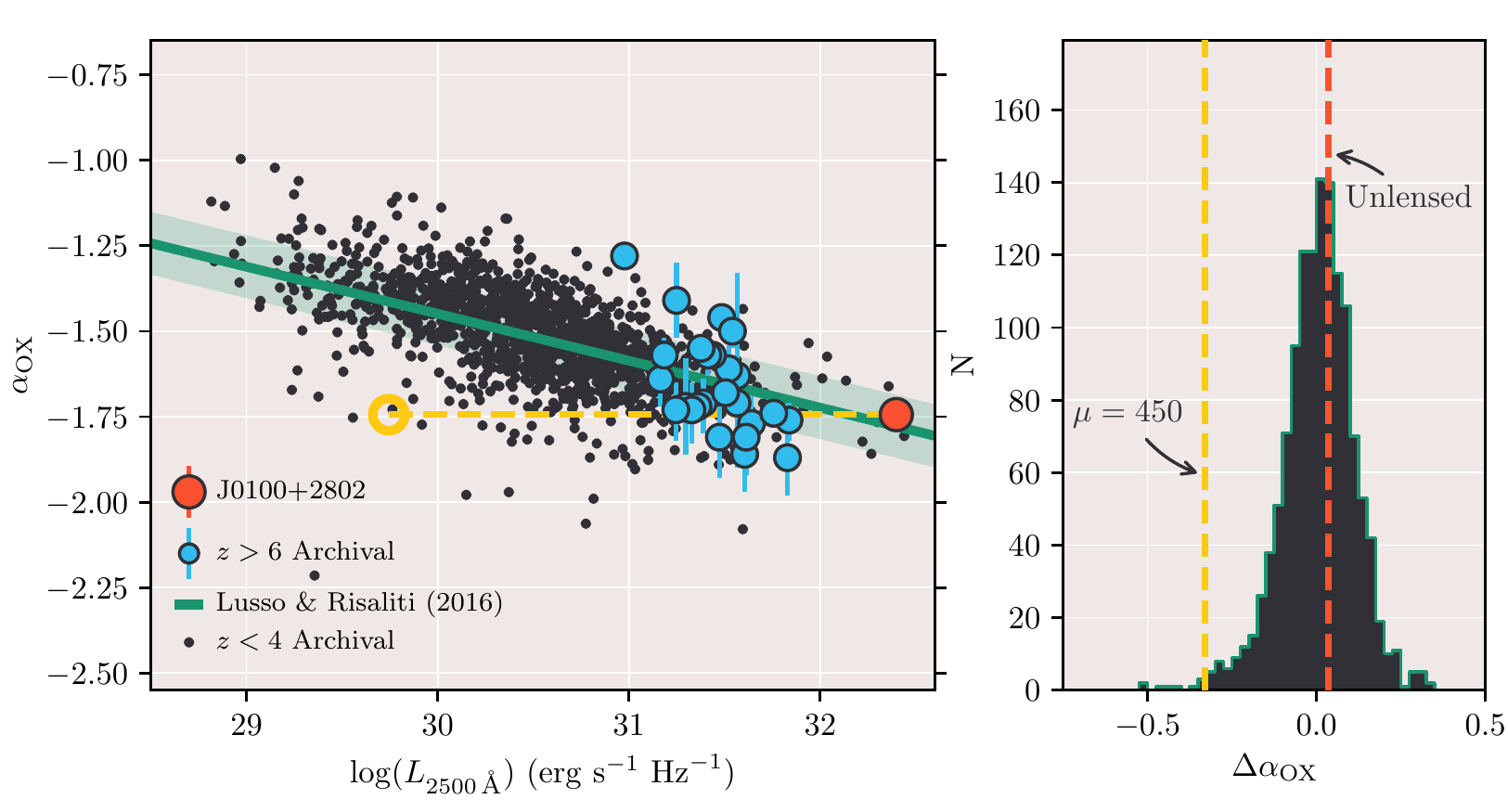}
      \caption{{\bf Left:} $\alpha_{OX}$ plotted against rest-frame UV monochromatic luminosity at 2500 \AA\ for \shortname\ (red if unlensed, yellow if magnified by $\mu=450$), a sample of $z\gtrsim6$ quasars with archival X-ray and UV data (blue), and low-redshift quasars (black). The predicted best relation from \citet{2016ApJ...819..154L} and its scatter are shown in green. {\bf Right:} The distribution of $\Delta \alpha_{\rm OX}$ values for the low-redshift sample. The positions of \shortname\ assuming no magnification (red) and the \citet{2020ApJ...891...64F} lensed model (yellow) are indicated by the vertical dashed lines. While \shortname\ is consistent with the predicted relation to within the scatter, it would be X-ray faint if it is magnified by $\mu\sim450$. Data sources are given in the text.
      }\label{fig:aox_plot}
\end{figure*}

\shortname\ was observed by \textit{XMM-Newton} for 65 ks in 2016, as previously reported by \citet{2017MNRAS.470.1587A}. We reprocessed these observations using the most current values of the \textit{XMM} calibrations and of the SAS analysis software, v19.0.0 \citep{2004ASPC..314..759G}. To minimize the effects of systematic processing choices, we used the \texttt{xmmextractor} routine to generate spectra from the EPIC camera observations, although we manually set the source  ($25^{\prime\prime}$ radius circle) and background regions ($90^{\prime\prime}-300^{\prime\prime}$ annulus with sources masked) to avoid contamination.

We note that \shortname\ was also observed by \textit{Chandra} in a short (14.8 ks), exploratory observation, as reported by \citet{2016ApJ...823L..37A}. However, that observation only obtained ${\sim}4\%$ of the total net counts obtained by the \textit{XMM} observations, so we do not include it in our spectral fitting. And, while \textit{Chandra} has a superior angular resolution to \textit{XMM}, it is not capable of resolving the small-separation ($0\farcs2$) images reported by \citet{2020ApJ...891...64F}. \texttt{BAYMAX}, the current state-of-the-art tool for identifying dual sources in \textit{Chandra} images, is of limited use for separations below $0\farcs3$, even if we had $50\times$ the number of \textit{Chandra} counts to work with \citep{2019ApJ...877...17F}.

The three EPIC spectra were fit simultaneously using the Python implementation of \texttt{XSPEC} \citep{1996ASPC..101...17A}. We binned our spectra to a minimum of only one count per bin, and so we found our best fits through the minimization of the modified $C$ statistic \citep{1979ApJ...228..939C, 1979ApJ...230..274W}. X-ray emission was modeled as an absorbed powerlaw, \texttt{phabs}$\times$\texttt{powerlaw}, with $n_H$ fixed to $6.21 \times 10^{20}\ {\rm cm}^{-2}$ \citep{2016A&A...594A.116H}. Our best-fits with $1\sigma$ uncertainties are $\Gamma=2.23^{+0.16}_{-0.13}$ and normalization $n=3.55^{+0.32}_{-0.31}\times10^{-6}$, corresponding to a broad-band, unabsorbed, rest-frame luminosity of $L_{2-10} = 4.63^{+0.62}_{-0.51}\times 10^{45} \ {\rm erg}\ {\rm s}^{-1}$ and monochromatic rest-frame luminosity at 2 keV of $L_{2\ {\rm keV}} = 7.1^{+1.7}_{-1.4} \times 10^{27}\ {\rm erg}\ {\rm s}^{-1}\ {\rm Hz}^{-1}$. These results are consistent with the previous analyses of the same \textit{XMM} data by \citeauthor{2019A&A...630A.118V} (\citeyear{2019A&A...630A.118V}; $L_{2-10} = 4.76^{+0.33}_{-0.31} \times 10^{45} \ {\rm erg}\ {\rm s}^{-1}$) and the (poorly-constrained) \textit{Chandra} results of \citeauthor{2016ApJ...823L..37A} (\citeyear{2016ApJ...823L..37A}; $L_{2-10} = 9.0^{+9.1}_{-4.5} \times 10^{45} \ {\rm erg}\ {\rm s}^{-1}$). \citeauthor{2017MNRAS.470.1587A} (\citeyear{2017MNRAS.470.1587A}) find a slightly lower luminosity from their analysis of the \textit{XMM} data ($L_{2-10} = 3.14^{+0.53}_{-0.48} \times 10^{45} \ {\rm erg}\ {\rm s}^{-1}$).

\begin{figure*}
\centering
\vspace{5mm}
   \includegraphics{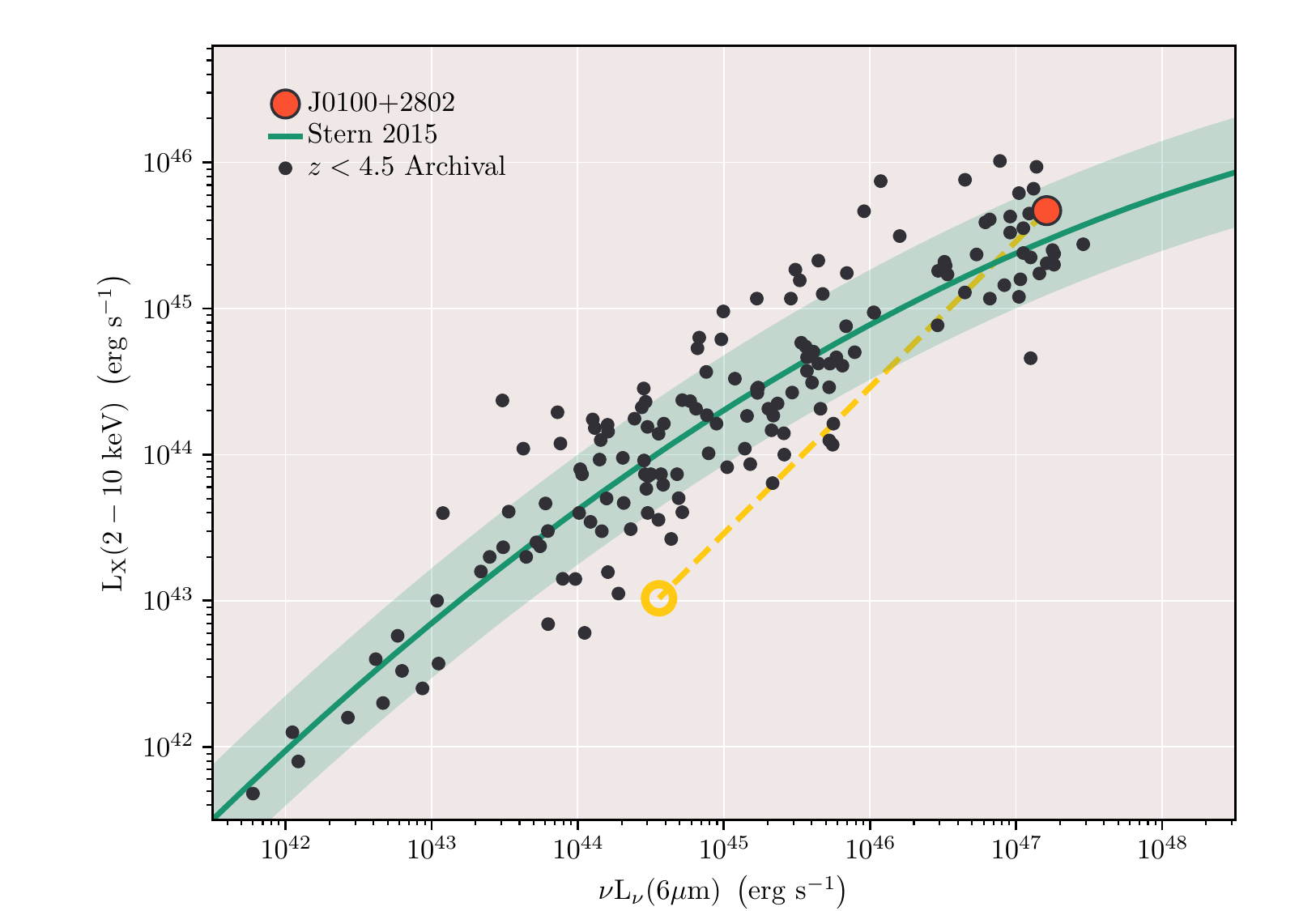}
      \caption{X-ray luminosity and mid-IR luminosity of \shortname\ (red) and a sample of lower redshift unlensed quasars (black, from the compilation of \citealt{2015ApJ...807..129S}). The scaling relation of \citet{2015ApJ...807..129S} is shown by the green line, with the shaded region corresponding to the standard deviation of offsets in the underlying sample, and the dashed yellow line shows the true position of \shortname\ if it were being lensed by $\mu\sim450$. Due to the bend in the X-ray-IR relation caused by the X-ray corona saturating at higher luminosities, lensing causes significant deviations from the nominal relation, as seen here.
      }\label{fig:stern_relation}
\end{figure*}

To enable comparisons to the expectation of the X-ray properties of \shortname, we also derive a monochromatic luminosity at rest-frame 2500 \AA\ ($L_{2500\, \textup{\footnotesize \AA}}$) and a rest-frame 6 $\mu$m luminosity ($\nu L_{6\ \mu{\rm m}}$). For the former, we use the fitted spectrum presented by \citet{2020ApJ...905...51S}, adopting $L_{2500\, \textup{\footnotesize \AA}} = 2.5 \times 10^{32}\ {\rm erg}\ {\rm s}^{-1}\ {\rm Hz}^{-1}$.
For $\nu L_{6\ \mu{\rm m}}$, we extrapolate from \textit{WISE} photometry, finding $\nu L_{6\ \mu{\rm m}}=1.6 \times 10^{47}\ {\rm erg}\ {\rm s}^{-1}$. While there is some uncertainty in this extrapolation, this value is consistent with the modeled mid-IR emission presented by \citet{2017MNRAS.470.1587A}.

\section{Comparison to Expectations}

One way to assess the expected strength of \shortname's X-ray emission is through the optical-to-X-ray flux ratio, $\alpha_{\rm OX}$\footnote{$\alpha_{\rm OX} \equiv 0.3838 \times \log\left(L_{2\ {\rm keV}} / L_{2500}\right)$, where $L_\nu$ is the monochromatic luminosity.}. X-ray emission has been observed to flatten as the rest-frame UV luminosity increases, thereby changing $\alpha_{\rm OX}$ with increasing UV luminosity. Magnification from lensing, however, would increase X-ray flux commensurate to the increase in UV brightness, thus leaving $\alpha_{\rm OX}$ unchanged. Here we rely on the best-fit of \citet{2016ApJ...819..154L}, who find
\begin{equation}
    \mathrm{log}({L}_{2\ {\rm{keV}}})= \gamma \mathrm{log}({L}_{2500})+\beta \pm \sigma,
\end{equation}
where $\gamma=0.642 \pm 0.015$ and $\beta={6.965}_{-0.465}^{+0.461}$ are fit parameters and $\sigma=0.24$ is the intrinsic scatter. We show this relation (projected into $\alpha_{\rm OX}$) and the observed values for \shortname\ in Figure \ref{fig:aox_plot}; the quasar's X-ray luminosity agrees with the expectation of its UV luminosity, assuming no magnification.

We also show two comparison samples in Figure \ref{fig:aox_plot}: a sample of $z>6$ quasars with measured X-ray properties (as compiled by \citealt{2021A&A...649A.133V}\footnote{These values are originally drawn from \citet{2019A&A...630A.118V}, \citet{2019ApJ...887..171C,2020ApJ...900..189C}, \citet{2021ApJ...908...53W}, and \citet{2021MNRAS.501.6208P}.}, excluding those values with only upper limits) and the low redshift sample used by \citet{2016ApJ...819..154L} to derive the $L_{2500}-L_{2\ {\rm keV}}$ relation. For the latter, we use the scaling relationship derived from the strictest cuts on data quality\footnote{$E(B-V) \leqslant 0.1$, X-ray ${\rm S}/{\rm N}>5$, and $1.9 \leqslant \Gamma \leqslant 2.8$}, but we show quasars filtered by less strict cuts\footnote{$E(B-V) \leqslant 0.1$, X-ray ${\rm S}/{\rm N}>5$, and $\Gamma \geqslant 1.6$}. We note the position where \shortname\ would intrinsically be were it observed with $\mu=450$, where it would be X-ray faint. Here, the offset from the predicted relation is $\Delta \alpha_{\rm OX} = -0.33$, corresponding to $3.5 \sigma$, based on the intrinsic scatter; by contrast, this is $\Delta \alpha_{\rm OX} = 0.03$ if \shortname\ is not magnified.

\begin{figure*}
\centering
\vspace{5mm}
   \includegraphics{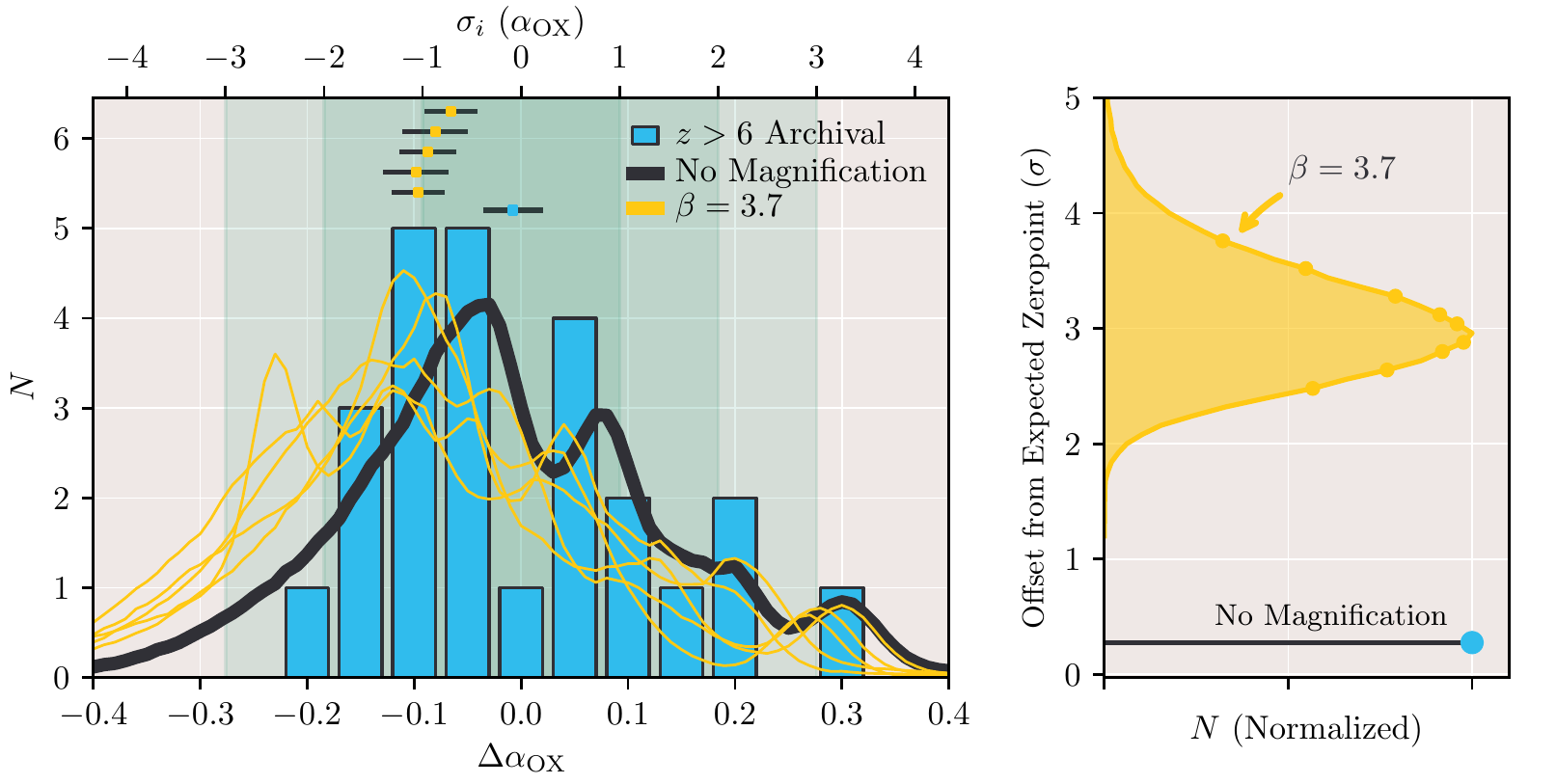}
      \caption{$\Delta \alpha_{OX}$ relative to the \citet{2016ApJ...819..154L} relation for the additional 25 $z>6$ quasars with measured X-ray properties (blue). When smoothed by a kernel of width equal to each quasar's individual uncertainties, the distribution is consistent with 0 (black). When the underlying magnification of each quasar is simulated 100,000 times following the \citet{2020ApJ...889...52P} $\beta=3.7$ model, this distribution shifts to negative values, with over half of the simulated quasars being fainter than expected by more than the \citet{2016ApJ...819..154L} intrinsic scatter. Five instances of this simulation are shown in yellow. The average value of $\Delta \alpha_{\rm OX}$ and corresponding standard error on the mean are shown above these distributions when assuming no magnification (blue with black errors) and for the five shown simulations (yellow with black errors). In the right panel, we show the results of the full suite of simulations. Quantified by the mean of $\Delta \alpha_{\rm OX}$ divided by the standard error in the mean, over 50\% of simulations have a $3\sigma$ or larger tension with the expectations of \citet{2016ApJ...819..154L}. Decile values are indicated by yellow markers.
      }\label{fig:delaox}
\end{figure*}

Another test of the expected X-ray luminosity comes from \citet{2015ApJ...807..129S}, who characterized a relationship between mid-IR and X-ray luminosities for $z<4.6$ AGN across almost six orders of magnitude, showing that the X-ray emission saturates at the highest mid-IR luminosities. As shown in Figure \ref{fig:stern_relation}, this saturation produces a distinct curve in the X-ray--mid-IR luminosity plane that, as noted by \citet{2020ApJ...895L..38S}, could potentially be used to identify lensed quasars. Lensing will magnify both luminosities equally, so that a lensed galaxy should rise away from the observed, curving relation. For \shortname, we have the opposite concern: its observed properties are almost exactly those predicted by the \citet{2015ApJ...807..129S} relation, and, were it lensed by $\mu{\sim}450$, it would have to be X-ray faint.

While \citet{2015ApJ...807..129S} do not report the scatter on their fit, we can nevertheless compute how significant a deviation \shortname\ would be from that relation if lensed. Using the same AGN sample \citet{2015ApJ...807..129S} used to compute the relation (excluding the obscured AGN from the \textit{NuSTAR} sample), we find a maximum negative offset of $\log(\Delta_{X} / {\rm erg}\ {\rm s}^{-1}) = -0.88$. In contrast, the nominal offset from this relation for \shortname, assuming $\mu=450$, is $\log(\Delta_{X} / {\rm erg}\ {\rm s}^{-1}) = -1.00$. Furthermore, the \citet{2015ApJ...807..129S} sample has a standard deviation of offsets of $\sigma=0.376\ \log({\rm erg}\ {\rm s}^{-1})$, meaning \shortname\ would be a $2.7\sigma$ offset. As such, the lensing scenario proposed by \citet{2020ApJ...891...64F} would not only require \shortname\ to be one of the most extremely-lensed quasars known, but it would also have to independently be an extreme outlier in X-ray faintness. 

\section{Is J0100+2802 X-ray Faint?}

As we have demonstrated above, if \shortname\ were lensed with a magnification of $\mu{\sim}450$, it would be extremely X-ray faint. Indeed, with $\Delta \alpha_{\rm OX}=-0.32$, it would be the second-most X-ray underluminous quasar known at $z>6$ and potentially even at $z>4$ \citep[][]{2021A&A...649A.133V}. Here we briefly discuss how likely this scenario is, independent of expectations from scaling relations.

There are three common AGN classifications that are each observed to be X-ray faint: Type 2 (obscured) AGN, broad absorption line (BAL) quasars, and weak emission-line quasars (WLQs, with Ly$\alpha$+\ion{N}{5} equivalent widths of ${\lesssim}10$ \AA). Type 2 AGN are not intrinsically X-ray weak relative to Type 1 AGN, but their soft X-ray flux is obscured by thick column densities ($n_H \gtrsim 10^{22}\ {\rm cm}^{-2}$, e.g., \citealt{2020MNRAS.491.5867R}), and they can be independently classified from their lack of broad emission lines in their UV/optical spectrum ($<1000\ {\rm km}\ {\rm s}^{-1}$, e.g., \citealt{2015ARA&A..53..365N}).
In contrast, BAL quasars are intrinsically X-ray weak, even at hard X-ray energies \citep{2014ApJ...794...70L}, and they are characterized by broad absorption lines.
We can rule these two classifications out, as deep ESI and X-Shooter spectra show that \shortname\ is neither type of quasar, with ${\rm FWHM}_{\rm Mg II} \sim 4000\ {\rm km}\ {\rm s}^{-1}$ and no broad absorption features \citep{2017ApJ...840...24E, 2020ApJ...905...51S}. 
\shortname\ is, however, a weak emission-line quasar, with a Ly$\alpha$+\ion{N}{5} equivalent width of ${\sim}10$ \AA\ \citep{2015Natur.518..512W}. 

Numerous studies have found WLQs are more likely to be X-ray faint \citep[e.g.,][]{2012ApJ...747...10W}, although the presence of weak lines does not guarantee X-ray weakness. One of the more common diagnostics for WLQs \citep[e.g.,][]{2018MNRAS.480.5184N} is the strength and velocity of the \ion{C}{4} line ($\lambda1549.06$ \AA), but \citet{2020ApJ...905...51S} were unable to model this line in their analysis of \shortname\ due to significant telluric contamination. However, there are other tracers for X-ray weakness in WLQs, notably the X-ray power law slope; for \shortname, $\Gamma=2.23^{+0.16}_{-0.13}$, which is a relatively soft value for high-redshift quasars \citep[e.g.][]{2019A&A...630A.118V}. \citet{2015ApJ...805..122L} find that X-ray weak WLQs have much harder X-ray spectra, with $\langle \Gamma \rangle = 1.16_{-0.32}^{+0.37}$, while normal-luminosity WLQs have $\langle \Gamma \rangle = 2.18\pm0.09$. Similarly, for a sample of 14 stacked $z\sim2$ X-ray faint WLQs, \citet{2018MNRAS.480.5184N} found $\Gamma = 1.19^{+0.56}_{-0.45}$. Furthermore, that work found rest-frame UV color was a predictor of X-ray weakness, being more common in redder quasars. \shortname, however, has $\alpha_\lambda=-1.55$ (where $f_\lambda \propto \lambda^{\alpha_\lambda}$), meaning its continuum is a typical color \citep[e.g.,][]{2001AJ....122..549V}. Thus, \shortname\ does not have optical properties that would lead to the expectation that it is X-ray faint.

We also directly compare the properties of \shortname\ to those of other quasars that are significant outliers in $\Delta \alpha_{\rm OX}$. In Figure \ref{fig:aox_plot} we show the distribution of $\Delta \alpha_{\rm OX}$ for the sample of \citet{2016ApJ...819..154L}, which has already been trimmed of BAL quasars; additionally, and as noted above, we also excluded sources with Galactic extinctions of $E(B-V) >0.1$ (following \citealt{2016ApJ...819..154L}), those with X-ray signal-to-noise ratios below 5, and those sources with X-ray spectral photon indices of $\Gamma < 1.6$, where the latter is indicative of obscuration \citep[e.g.,][]{2017ApJS..233...17R}. Although the unlensed value of \shortname\ is consistent with the expectations of \citet{2016ApJ...819..154L}, for the $\mu=450$ scenario, \shortname\ resides on the extreme tail of the $\Delta\alpha_{\rm OX}$ distribution. Similarly, if lensed, \shortname\ would be fainter than the mid-IR prediction at a level unmatched in the \citet{2015ApJ...807..129S} sample.

The \citet{2020ApJ...891...64F} lensing hypothesis requires \shortname\ to be at the extremes of quasar magnification \citep[cf., e.g.,][]{2018arXiv180705434G}, while the analysis here shows that it would also have to be a significant outlier in X-ray faintness, despite displaying neither the optical nor X-ray characteristics expected for such an object. While, in the infiniteness of the Universe, it is possible for a quasar with a $3\sigma$ underluminosity to also be lensed by a similarly extreme magnification, these two probabilities are independent, and so Occam's razor leads us to the much simpler conclusion: \shortname\ is only marginally magnified, at most.

\section{Constraints from the Broader Population}

Even if we accept enough uncertainty to allow for \shortname\ to be an order of magnitude fainter than predicted, there is still one further test. If \shortname\ were magnified by $\mu{\sim}450$, \citet{2020ApJ...889...52P} argue that we would expect a significant fraction of all known $z>6$ quasars to be lensed as well. \citet{2020ApJ...889...52P} find that the detection of a source with such an extreme magnification requires $\beta$, the bright-end slope of the luminosity function  -- such that $\Phi(L) \propto L^{-\beta}$ -- must be $\beta \geq 3.7$. From this value we can derive $P(\mu)$, the probability of an individual quasar being lensed to a certain magnification \citep{2019ApJ...870L..12P}. As such, there is another observable test of the \citet{2020ApJ...891...64F} hypothesis: are the remaining $z>6$ quasars consistent with being magnified at the level predicted for $\beta=3.7$?

To test this proposition, we consider the $\Delta\alpha_{\rm OX}$ of all the $z>6$ quasars plotted in Figure \ref{fig:aox_plot}. To standardize the analysis, we compute $L_{2500}$ for these objects directly from their measured ${\rm M}_{1450}$ values, assuming $L_\nu \propto \nu^{-\alpha_\nu}$ and $\alpha_\nu=0.3$ \citep[e.g.,][]{2016A&A...585A..87S}, and we compare $\alpha_{\rm OX}$ values to those predicted by \citet{2016ApJ...819..154L}. The individual values of $\Delta\alpha_{\rm OX}$ are shown by the blue histogram in Figure \ref{fig:delaox}. To account for uncertainties, we smooth this distribution with a Gaussian kernel sized to each quasar's $\alpha_{\rm OX}$ uncertainty, which is shown in the solid black line. The mean and standard error on the mean for this distribution are $-0.008$ and $\pm0.028$, respectively.

For the $\beta=3.7$ case, we simulate what the values of $\Delta\alpha_{\rm OX}$ would be if each quasar was unmagnified for a randomly-drawn value of $P(\mu)$. Magnification probabilities are derived from \citet{2020ApJ...889...52P}, who report ${\displaystyle P_{51}(\mu) = \prod_{i=1}^{51} (1 - P_i({\geq}\mu))}$, the cumulative probability that at least one quasar in a sample of 51 is magnified by $\mu$. Here, we assume all quasars have the same inherent lensing probabilities. As this distribution stops at $\mu=10$, we linearly extrapolate in $\log\left(P({\geq}\mu)\right) - \log(\mu)$ space, in keeping with the distributions shown for $\beta=3.6$ and $\beta=2.6$ by \citet{2019ApJ...870L..12P}. To reduce the effects of this extrapolation on our results, we conservatively decrease the projected slope by 10\%. 

We then simulated new distributions of $\Delta\alpha_{\rm OX}$ by probabilistically assigning each quasar an assumed magnification and then computing that quasar's lensing-corrected offset. In more than 50\% of our simulations, the average value of $\Delta\alpha_{\rm OX}$ exceeded a  $3\sigma$ negative offset from 0, quantified by the standard error on the mean; the full distribution of simulation results is shown in the right panel of Figure \ref{fig:delaox}. We also show five individual realizations of the simulation on the left panel, again smoothed by measurement uncertainties. 

The implication of these results is that, if \shortname\ is lensed by $\mu{\sim}450$, $z>6$ quasars must behave fundamentally differently at X-ray energies than at lower redshifts. However, \citet{2020A&A...642A.150L} have shown that quasars show no evolution in their X-ray to UV relation up to at least $z{\sim}5$, while \citet{2017A&A...603A.128N} report that there is no evolution in the basic X-ray spectral properties up to $z{\sim}6$. Furthermore, \citet{2014ApJ...790..145D} find no evolution in the emission line properties (flux ratios and equivalent widths) of $z>6$ quasars compared to their low-redshift counterparts, while \citet{2020MNRAS.492..719T} find that these emission lines (\ion{Mg}{2} and \ion{C}{4}) are tracers of the underlying $L_{2500}-L_{2\ {\rm keV}}$ relation. And, while there is some evidence that WLQs are more common at higher redshifts \citep{2016ApJS..227...11B}, the X-ray properties of our sample are consistent with not being X-ray faint \citep{2019A&A...630A.118V}. As such, the large deviation from the $\alpha_{\rm OX}$ relation required for \shortname\ to be strongly-lensed is in tension with our current understanding of early AGN.

\section{Summary}

In this letter, we have examined the claims by \citet{2020ApJ...891...64F} that \shortname\ is lensed with a magnification of $\mu{\sim}450$ through the perspective of the quasar's X-ray properties. In particular, we focus on how the source's X-ray luminosity relates to its rest-frame UV and IR luminosities. We find that, for \shortname\ to be magnified at this level, it would have to be an extremely yet uncharacteristically intrinsically-faint X-ray source. In addition, the implication posed by \citet{2020ApJ...889...52P} that many $z\gtrsim6$ quasars are lensed is incompatible with broader trends in AGN evolution established in the wider literature. The quantification of our analysis includes:
\begin{itemize}
    \item If lensed by $\mu{\sim}450$, \shortname\ is offset from the \citet{2016ApJ...819..154L} $L_{2500}-\alpha_{\rm OX}$ relation by $\Delta \alpha_{\rm OX} = -0.32$, which corresponds to $3.5\sigma$, based on the intrinsic scatter. The spectral characteristics of \shortname\ are not consistent with those of typical quasars that are X-ray faint.
    \item If lensed by $\mu{\sim}450$, \shortname\ is offset from the \citet{2015ApJ...807..129S} $L_X - \nu L_{6\ \mu{\rm m}}$ relation by $\log(L_{2 - 10\ {\rm keV}}) = -1.00$ ($2.7\sigma$), which would be larger than any other quasar in the \citet{2015ApJ...807..129S} sample ($N=155$).
    \item If the population of $z>6$ quasars is lensed following the predictions of \citet{2020ApJ...889...52P}, which are needed to produce a magnification of $\mu=450$, then these quasars would be in tension with the \citet{2016ApJ...819..154L} $L_{2500}-\alpha_{\rm OX}$ relation for lower-redshift quasars at a $\gtrsim 3\sigma$ level, despite extensive evidence that the central engines of high-redshift quasars are not significantly different from those of low-redshift quasars.
\end{itemize}

Further insight into the nature of \shortname\ should come with the launch of the \textit{JWST}, as this quasar is the target of guaranteed-time observations, including a spatial investigation of the host galaxy with NIRSpec-IFU \citep{2017jwst.prop.1218F}. While awaiting those data, we demonstrate here that X-ray observations remain a powerful tool for studying the early Universe.

\vspace{2mm}
{\small \noindent The authors thank the anonymous referee for their constructive feedback in the preparation of this article and Fabio Pacucci for providing the lensing probabilities associated with the $\beta=3.7$ model. The work of T.C. and D.S. was carried out at the Jet Propulsion Laboratory, California Institute of Technology, under a contract with the National Aeronautics and Space Administration (80NM0018D0004). T.C.'s research was supported by an appointment to the NASA Postdoctoral Program at the Jet Propulsion Laboratory, California Institute of Technology, administered by Universities Space Research Association under contract with NASA.

Based on observations obtained with XMM-Newton, an ESA science mission with instruments and contributions directly funded by ESA Member States and NASA.}

\vspace{-1mm}
\facility{XMM}
\software{PyFITS \citep{1999ASPC..172..483B},
          SAS \citep{2004ASPC..314..759G},
          XSPEC \citep{1996ASPC..101...17A}}

\textcopyright\ 2021. All rights reserved.

\vspace{-1mm}
\bibliography{bibliography}{}

\end{document}